\documentclass[a4paper,12pt]{amsart}
\usepackage{txfonts,epic}

\def\?{\mskip2.3mu}
\def\rp{r'\!}
\def\xr{X}
\begin{document}
\title[On Zamolodchikov's Conjecture]
{On Zamolodchikov's Periodicity Conjecture
for Y-systems}
\author[A.Yu. Volkov]{Alexandre Yu. Volkov}
\address{Dienst Theoretische Natuurkunde,
Vrije Universiteit Brussel,
Pleinlaan 2, 1050 Brussel, Belgium}
\address{Instituts Solvay,
Campus Plaine C.P. 231,
Boulevard du Triomphe,
1050 Bruxelles, Belgium}
\begin{abstract}
I prove Zamolodchikov's periodicity conjecture for
type~$A$ with both ranks arbitrary.
\end{abstract}
\maketitle
\section*{Introduction}
Following Alexei Zamolodchikov \cite{z}, one can
associate to any pair of indecomposable Cartan
matrices of finite type the so called Y-system of
algebraic equations, which reads
\begin{gather}\label{orig}
   Y_{i\,j+1\?k}Y_{i\,j-1\?k}
   =\frac{{\displaystyle\prod_{i'\neq\?i}}
   \,\left(1+Y_{i'\!j\?k}\right)^{-a_{ii'}}}
   {{\displaystyle\prod_{k'\neq\?k}}\,
   \left(1+1/Y_{ij\?k'}
   \right)^{-a'_{\smash{kk'}}}}\\[1ex]\nonumber
   \;\;i,j,\?k\in\mathbb Z\qquad
   1\leq i\leq r\qquad
   1\leq k\leq \rp\;,
\end{gather}
where $(a_{ii'})$ and $(a'_{\smash{kk'}})$ are those
Cartan matrices, and~$r$ and~$\rp$ are their ranks.
Then the periodicity conjecture asserts that all
solutions to this system are periodic in~$j$, with
period equal to twice the sum of the respective dual
Coxeter numbers.
 
Until now this conjecture has only been proved in the
case when one of the ranks equals~1 \cite{fs,gt,fz}.
In this paper, we will take the next logical step and
prove the case when the two Cartan matrices are of
type $A$ of arbitrary ranks $r$ and $\rp$.

Recall that Cartan matrices of type $A$ are
tridiagonal, with twos on the diagonal and minus ones
on the sub- and sup-diagonals. Thus, for $1<i<r$ and
$1<k<\rp$ we have
\begin{equation*}
   Y_{i\,j+1\?k}\?Y_{i\,j-1\?k}=\frac
   {\left(1+Y_{i+1j\?k}\right)
   \left(1+Y_{i-1j\?k}\right)}
   {\left(1+1/Y_{ij\?k+1}\right)
   \left(1+1/Y_{ij\?k-1}\right)}\;,
\end{equation*}
while at the boundaries one or two factors in the
right hand side are absent---for instance,
\begin{gather*}
   Y_{1\?j+1\?k}Y_{1\?j-1\?k}
   =\frac{1+Y_{2j\?k}}
   {\left(1+1/Y_{1j\?k+1}\right)
   \left(1+1/Y_{1j\?k-1}\right)}\\[1ex]
   Y_{1\?j+1\?1}Y_{1\?j-1\?1}=\frac
   {1+Y_{2j\?1}}{1+1/Y_{1j\?2}}\;.
\end{gather*}
However, they can be added back if we introduce
fictitious boundary variables and set them equal
zero or infinity, as appropriate: 
$Y_{0j\?k}=Y_{r+1j\?k}=0$,
$Y_{ij\,0}=Y_{ij\?\rp+1}=\infty$.
Note that
$Y_{0j\,0}$, $Y_{0j\?\rp+1}$, $Y_{r+1j\,0}$ and
$Y_{r+1j\?\rp+1}$ are thus ill-defined, but they
never appear in the right hand side anyway.

Note also that our system consists of two
completely decoupled identical sub\-systems---one
involving variables $Y_{ij\?k}$ with $i+j+k$ even,
and the other with $i+j+k$ odd. We will, therefore,
simply discard the second subsystem and assume that
the $Y_{ij\?k}$ are only defined for $i+j+k$ even.

Finally, recall that the dual Coxeter number for
type $A$ equals rank plus one. Thus, we have to
prove the following.\\[1ex]
\noindent {\bf Theorem} (Zamolodchikov's conjecture,
case $AA$).
\em All regular solutions (that is, solutions
avoiding $0$, $-1$ and $\infty$ everywhere except
the boundaries) to the Y-system
\begin{gather}\label{ysys}
   Y_{i\,j+1\?k}\?Y_{i\,j-1\?k}=\frac
   {\left(1+Y_{i+1j\?k}\right)
   \left(1+Y_{i-1j\?k}\right)}
   {\left(1+1/Y_{ij\?k+1}\right)
   \left(1+1/Y_{ij\?k-1}\right)}\\[1ex]
   \nonumber 1\leq i\leq r\qquad
   1\leq k\leq \rp\qquad
   i+j+k\;\;\text{odd}
\end{gather}
with the `free boundary conditions'
\begin{equation}\label{bc}
   Y_{0j\?k}=Y_{r+1j\?k}=0\qquad\qquad
   Y_{ij\,0}=Y_{ij\,\rp+1}=\infty
\end{equation}
are $2(r+\rp+2)$-periodic in $j$:
\begin{equation}\label{per}
   Y_{i,j+2(r+\rp+2),\?k}=Y_{ij\?k}\;.
\end{equation}

\em We prove this by producing a manifestly
periodic formula for the general solution. This
formula involves $r+\rp+2$ arbitrary points of
the $r$-dimensional projective space,
\[
   x_n\in\mathbb{CP}^r\qquad\qquad
   x_{n+r+\rp+2}=x_n\;,
\]
and reads
\begin{equation}\label{y}
   Y_{ij\?k}=-\,\xr\!\underbrace{_
   {x_{a+1}+\dotsb+x_{b-1}}}_{i-1}{\!}_
   +\!\underbrace{_
   {x_{c+1}+\dotsb+x_{d-1}}}_{r-i}
   (x_a,x_b,x_c,x_d)\;,
\end{equation}
where
\begin{xalignat*}{2}
   a&=\frac{-i-j-k}2&
   b&=a+i=\frac{i-j-k}2\\[1ex]
   c&=b+k=\frac{i-j+k}2&
   d&=c+r+1-i=\frac{-i-j+k}2+r+1\;,
\end{xalignat*}
and $\xr$ is a certain $(r+3)$-point projective
invariant to be defined in Section~1.

As you can see, periodicity is indeed manifest, and
so we shall be done once we check that a) the
proposed solution does actually solve the
Y-system~(\ref{ysys}), and b) that it is indeed
a general solution. This is done in Sections~2
and~3.

I wish to thank L. Faddeev, I. Krichever, I. Loris,
A. Szenes and especially A. Alekseev for helpful
discussions.
\section{Cross-ratio}
In this section we recall some textbook facts about
projective geometry and define the projective
invariant used in formula~(\ref{y}).

Recall that the projective space $\mathbb{CP}^r$
is the set of all lines through the origin of
$\mathbb C^{r+1}$, or in other words, it is the
quotient of $\mathbb C^{r+1}\backslash\{0\}$ by the
equivalence relation that two vectors are equivalent
iff they are collinear. For a vector
$\vec x\in\mathbb C^{r+1}\backslash\{0\}$ we denote
the corresponding element (point) of $\mathbb{CP}^r$
by $[\vec x\,]$ or simply~$x$, and say that the
former is a representative vector of the latter.
Points are said to be projectively independent if
their representative vectors are linearly
independent.

The projective span $x_1+\dotsb+x_N$ of two or more
points $x_n\in\mathbb{CP}^r$ is the set of all points
whose representative vectors lie in the linear span
of the vectors~$\vec x_n$. Projective spans of $r$
projectively independent points are called
hyperplanes.

To any invertible linear transformation $T$ of
$\mathbb C^{r+1}$ we associate the invertible map~$t$
of $\mathbb{CP}^r$ onto itself defined by
$t\left([\vec x\,]\right)=\left[T\vec x\,\right]$.
Such maps are called projective transformations, and
they map projective spans to projective spans:
$t\?(x_1+\dotsb+x_N)=t\?(x_1)+\dotsb+t\?(x_N)$.

We say $r+2$ points of $\mathbb{CP}^r$ to be in
general position, or to form a projective frame, if
no $r+1$ of them lie in the same hyperplane, or in
other words, if no $r+1$ of their representative
vectors are linearly dependent. Such frames are
akin to vector bases in that for any two of them
there is a unique projective transformation taking
one to the other.

For instance, for $r=1$ any three distinct points
form a projective frame. Thus, they can be taken
into any other three by a unique projective
transformation, but this is not the case for four
points. Four points in $\mathbb{CP}^1$ (of which at
least three are distinct) have a numeric
projective invariant, which is called the
cross-ratio, and is defined by
\[
   \xr\?(x_1,x_2,x_3,x_4)
   =\frac{\vec x_1\wedge\vec x_2}
   {\vec x_1\wedge\vec x_4}\,
   \frac{\vec x_3\wedge\vec x_4}
   {\vec x_3\wedge\vec x_2}\;.
\]
Note that division of bivectors makes proper sense
here because they all are multiples of one and the
same bivector. Explicitly, if~$x^{\?l}_n$ ($l=1,2$)
are coordinates of vectors~$\vec x_n$ relative to
some basis $\vec e_1,\vec e_2\in\mathbb C^2$, and
$|x_m\,x_n|$ denotes the determinant of the
corresponding coordinate matrix,
\[
   |\vec x_m\,\vec x_n|
   =\det\left(\begin{matrix}x^1_m&x^1_n\\[1ex]
   x^2_m&x^2_n\end{matrix}\?\right)
\]
then $\vec x_m\wedge\vec x_n
=|\vec x_m\,\vec x_n|\,\vec e_1\wedge\vec e_2$,
and hence
\[
   \xr\?(x_1,x_2,x_3,x_4)=\frac
   {|\vec x_1\,\vec x_2|\?|\vec x_3\,\vec x_4|}
   {|\vec x_1\,\vec x_4|\?|\vec x_3\,\vec x_2|}
   =\frac{(z_1-z_2)(z_3-z_4)}
   {(z_1-z_4)(z_3-z_2)}\;,
\]
where $z_n=x^1_n/x^2_n\in\mathbb C\cup\{\infty\}$.

The last expression makes it particularly easy to
check how the cross-ratio changes under permutations
of the four points involved. It turns out that
24 such permutations yield only 6 different values of
the cross-ratio: if one of them equals~$\omega$, then
the other five are $1/\omega$, $1-\omega$,
$1/(1-\omega)$, $1-1/\omega$ and $1/(1-1/\omega)$. In
particular, we have
\begin{gather*}
   \xr\?(x_1,x_3,x_2,x_4)
   =1-\xr\?(x_1,x_2,x_3,x_4)\\[1ex]
   \xr\?(x_1,x_2,x_4,x_3)
   =\frac1{1-1/\xr\?
   (x_1,x_2,x_3,x_4)}\;,
\end{gather*}
which suitably explains why the Y-system involves
$Y$, $1+Y$ and $1/(1+1/Y)$ at the same time.

Back to the general case, we are now ready to define
the invariant used in the general solution
formula~(\ref{y}). Consider $r+3$ points
$x_1,\dotsc,x_{r+3}\in\mathbb{CP}^r$ such that any
three of the first four plus the remaining $r-1$ of
them are in general position. Let $V$ be the subspace
of $\mathbb C^{r+1}$ spanned by the representative
vectors of those last $r-1$ points
$x_5,\dotsc,x_{r+3}$, and $\pi$ the canonical
projection $\mathbb C^{r+1}\rightarrow
\mathbb C^{r+1}/V\simeq\mathbb C^2$.
Then the multi-dimensional cross-ratio is defined by
\begin{equation}
   \xr_{x_5+\dotsb+x_{r+3}}(x_1,x_2,x_3,x_4)
   =\xr\?(x'_1,x'_2,x'_3,x'_4)\;,
\end{equation}
where $x'_n=[\pi(\vec x_n)]\in\mathbb{CP}^1$.

Note two things. First, that this generalized
cross-ratio obviously has the same behaviour under
permutation of the points $x_1,\dots,x_4$ as did the
original cross-ratio. In particular, we again have
\begin{equation}\label{perm}\begin{aligned}
   \xr_{x_5+\dotsb+x_{r+3}}(x_1,x_3,x_2,x_4)&=1-
   \xr_{x_5+\dotsb+x_{r+3}}(x_1,x_2,x_3,x_4)\\[1ex]
   \xr_{x_5+\dotsb+x_{r+3}}(x_1,x_2,x_4,x_3)
   &=\frac1{1-1/\xr_{x_5+\dotsb+x_{r+3}}
   (x_1,x_2,x_3,x_4)}\;.
\end{aligned}\end{equation}

Second, that, in terms of wedge products and
determinants, the above defition becomes
\begin{multline}\label{xrr}
   \xr_{x_5+\dotsb+x_{r+3}}(x_1,x_2,x_3,x_4)
   =\frac{\vec x_1\wedge\vec x_2\wedge\vec x_5
   \wedge\dotsb\wedge\vec x_{r+3}}
   {\vec x_1\wedge\vec x_4\wedge\vec x_5
   \wedge\dotsb\wedge\vec x_{r+3}}\\[1ex]
   \times\frac{\vec x_3\wedge\vec x_4\wedge\vec x_5
   \wedge\dotsb\wedge\vec x_{r+3}}
   {\vec x_3\wedge\vec x_2\wedge\vec x_5
   \wedge\dotsb\wedge\vec x_{r+3}}
   =\frac{|\vec x_1\,\vec x_2\,\vec x_5\dotsc
   \vec x_{r+3}|\?
   |\vec x_3\,\vec x_4\,\vec x_5\dotsc
   \vec x_{r+3}|}
   {|\vec x_1\,\vec x_4\,\vec x_5\dotsc
   \vec x_{r+3}|\?
   |\vec x_3\,\vec x_2\,\vec x_5\dotsc
   \vec x_{r+3}|}\;.
\end{multline}
It is the latter expression that we mostly use in
what follows.
\section{Proof, part $(\mathrm a)$}
Here we check that the $Y$'s given by
formula~(\ref{y}) do indeed satisfy the
Y-system (\ref{ysys}).

First, we use eq.~(\ref{xrr}) to restate that formula
in a fully explicit form. This time it involves
$r+\rp+2$ arbitrary $(r+1)$-vectors
\[
   \vec x_n\in\mathbb C^{r+1}\qquad\qquad
   \vec x_{n+r+\rp+2}=\vec x_n\;,
\]
and reads
\begin{equation}\label{yc}
   Y_{ij\?k}=\frac
   {|\vec x_a\dotsc\vec x_b\,
   \vec x_{c+1}\dotsc\vec x_{d-1}|\?
   |\vec x_{a+1}\dotsc\vec x_{b-1}\,
   \vec x_c\dotsc\vec x_d|}
   {|\vec x_a\dotsc\vec x_{b-1}\,
   \vec x_{c+1}\dotsc\vec x_d|\?
   |\vec x_{a+1}\dotsc\vec x_b\,
   \vec x_c\dotsc\vec x_{d-1}|}\;,
\end{equation}
where, as before,
\begin{xalignat*}{2}
   a&=\frac{-i-j-k}2&
   b&=a+i=\frac{i-j-k}2\\
   c&=b+k=\frac{i-j+k}2&
   d&=c+r+1-i=\frac{-i-j+k}2+r+1\;.
\end{xalignat*}
Note that the minus before the right hand side is
gone because the order of columns here is different
from that in the original definition~(\ref{xrr}).
Note also that the inequalities $0<i<r+1$ and
$0<k<\rp+1$ are exactly equivalent to
$a<b<c<d<a+r+\rp+2$. This guarantees that
formula~(\ref{yc}) makes sense for all~$i$ and~$k$
in the range, and none of the determinants involved
is identically zero.

Now, denote the upper left determinant in the right
hand side by $\Delta_{\?i+1j\?k}$. Then, clearly,
the remaining three are $\Delta_{\?i-1j\?k}$,
$\Delta_{\?ij\?k+1}$ and $\Delta_{\?ij\?k-1}$,
\begin{equation}\label{ca}
   Y_{ij\?k}=\frac
   {\Delta_{\?i+1j\?k}\,\Delta_{\?i-1j\?k}}
   {\Delta_{\?ij\?k+1}\,\Delta_{\?ij\?k-1}}\;,
\end{equation}
and likewise
\[
   \Delta_{\?i\?j+1\?k}
   =|\vec x_a\dotsc\vec x_{b-1}\,
   \vec x_c\dotsc\vec x_{d-1}|\qquad
   \Delta_{\?i\?j-1\?k}
   =|\vec x_{a+1}\dotsc\vec x_b\,
   \vec x_{c+1}\dotsc\vec x_d|\;,
\]
Hence, by eqs.~(\ref{perm}),
\[
   1+Y_{ij\?k}=\frac
   {\Delta_{\?i\?j+1\?k}\,\Delta_{\?i\?j-1\?k}}
   {\Delta_{\?ij\?k+1}\,\Delta_{\?ij\?k-1}}
   \qquad\qquad
   \frac1{1+1/Y_{ij\?k}}=\frac
   {\Delta_{\?i+1j\?k}\,\Delta_{\?i-1j\?k}}
   {\Delta_{\?i\?j+1\?k}\,\Delta_{\?i\?j-1\?k}}\;,
\]
and therefore, indeed,
\begin{multline*}
   Y_{i\?j+1\?k}\?Y_{i\?j-1\?k}=\frac
   {\Delta_{\?i+1j+1\?k}\,\Delta_{\?i-1j+1\?k}}
   {\Delta_{\?i\?j+1\?k+1}\,\Delta_{\?i\?j+1\?k-1}}\,
   \frac
   {\Delta_{\?i+1j-1\?k}\,\Delta_{\?i-1j-1\?k}}
   {\Delta_{\?i\?j-1\?k+1}\,\Delta_{\?i\?j-1\?k-1}}
   \\[1ex]=\frac
   {\Delta_{\?i+1j+1\?k}\,\Delta_{\?i+1j-1\?k}}
   {\Delta_{\?i+1j\?k+1}\,\Delta_{\?i+1j\?k-1}}\,
   \frac{\Delta_{\?i-1j+1\?k}\,\Delta_{\?i-1j-1\?k}}
   {\Delta_{\?i-1j\?k+1}\,\Delta_{\?i-1j\?k-1}}\,
   \\[1ex]\times
   \frac{\Delta_{\?i+1j\?k+1}\,\Delta_{\?i-1j\?k+1}}
   {\Delta_{\?i\?j+1\?k+1}\,\Delta_{\?i\?j-1\?k+1}}\,
   \frac{\Delta_{\?i+1j\?k-1}\,\Delta_{\?i-1j\?k-1}}
   {\Delta_{\?i\?j+1\?k-1}\,\Delta_{\?i\?j-1\?k-1}}
   =\frac{\left(1+Y_{i+1j\?k}\right)
   \left(1+Y_{i-1j\?k}\right)}
   {\left(1+1/Y_{ij\?k+1}\right)
   \left(1+1/Y_{ij\?k-1}\right)}\;,
\end{multline*}
at least for $1<i<r$ and $1<k<\rp$. The boundary
cases are checked separately using the fact that, by
construction,
\begin{xalignat*}{2}
   \Delta_{\?0j\?k}&=\Delta_{\?0j-1\?k-1}&
   \Delta_{\?r+1j\?k}&=\Delta_{\?r+1j-1\?k+1}\\
   \Delta_{\?ij\,0}&=\Delta_{\?i+1j-1\?0}&
   \Delta_{\?ij\,\rp+1}
   &=\Delta_{\?i-1j-1\?\rp+1}\;.
\end{xalignat*}
Thus, for instance,
\begin{multline*}
   Y_{1\?j+1\?k}Y_{1\?j-1\?k}=\frac
   {\Delta_{\?2j+1\?k}\,\Delta_{\?0j+1\?k}}
   {\Delta_{\?1j+1\?k+1}\,\Delta_{\?1j+1\?k-1}}\,
   \frac{\Delta_{\?2j-1\?k}\,\Delta_{\?0j-1\?k}}
   {\Delta_{\?1j-1\?k+1}\,\Delta_{\?1j-1\?k-1}}
   =\frac{\Delta_{\?2j+1\?k}\,\Delta_{\?2j-1\?k}}
   {\Delta_{\?2j\?k+1}\,\Delta_{\?2j\?k-1}}
   \\[1ex]\times
   \frac{\Delta_{\?2j\?k+1}\,\Delta_{\?0j\?k+1}}
   {\Delta_{\?1j+1\?k+1}\,\Delta_{\?1j-1\?k+1}}\,
   \frac{\Delta_{\?2j\?k-1}\,\Delta_{\?0j\?k-1}}
   {\Delta_{\?1j+1\?k-1}\,\Delta_{\?1j-1\?k-1}}
   =\frac{1+Y_{2j\?k}}
   {\left(1+1/Y_{1j\?k+1}\right)
   \left(1+1/Y_{1j\?k-1}\right)}
\end{multline*}
and
\begin{multline*}
   Y_{1\?j+1\?1}Y_{1\?j-1\?1}=\frac
   {\Delta_{\?2j+1\?1}\,\Delta_{\?0j+1\?1}}
   {\Delta_{\?1j+1\?2}\,\Delta_{\?1j+1\?0}}\,
   \frac{\Delta_{\?2j-1\?1}\,\Delta_{\?0j-1\?1}}
   {\Delta_{\?1j-1\?2}\,\Delta_{\?1j-1\?0}}\\[1ex]
   =\frac{\Delta_{\?2j+1\?1}\,\Delta_{\?2j-1\?1}}
   {\Delta_{\?2j\,2}\,\Delta_{\?2j\,0}}\,
   \frac{\Delta_{\?2j\,2}\,\Delta_{\?0j\,2}}
   {\Delta_{\?1j+1\?2}\,\Delta_{\?1j-1\?2}}
   =\frac{1+Y_{2j\?1}}{1+1/Y_{1j\,2}}\;,
\end{multline*}
as it should be (see Introduction). The remaining
six boundary cases are checked similarly, and this
completes this part of the proof.

A remark is in order here. In our approach
determinants $\Delta_{\?ij\?k}$ play a secondary
part, but, as it turns out, they satisfy their own
system of algebraic equations rather similar to
the Y-system. Indeed, note that either of the
relations~(\ref{perm}) is equivalent to the
classical bilinear determinant identity
\begin{multline*}
   \qquad|\vec x_1\,\vec x_3\,\vec x_5\dotsc
   \vec x_{r+3}|\?
   |\vec x_2\,\vec x_4\,\vec x_5\dotsc
   \vec x_{r+3}|\\
   =|\vec x_1\,\vec x_2\,\vec x_5\dotsc
   \vec x_{r+3}|\?
   |\vec x_3\,\vec x_4\,\vec x_5\dotsc
   \vec x_{r+3}|\\
   +|\vec x_1\,\vec x_4\,\vec x_5\dotsc
   \vec x_{r+3}|\?
   |\vec x_2\,\vec x_3\,\vec x_5\dotsc
   \vec x_{r+3}|\;,\qquad
\end{multline*}
also known as the Pl\"ucker relation. Hence, the
$\Delta$'s satisfy the lattice system
\begin{equation*}
   \Delta_{\?i\?j+1\?k}\,\Delta_{\?i\?j-1\?k}
   =\Delta_{\?i+1j\?k}\,\Delta_{\?i-1j\?k}
   +\Delta_{\?ij\,k+1}\,\Delta_{\?ij\,k-1}\;,
\end{equation*}
which is well known in Lattice Soliton Theory as the
3D Hirota equation. Since the Y-systems and this
Hirota equation are related to each other by a simple
change of variables~(\ref{ca}), Zamolodchikov's
conjecture could be easily reformulated in terms of
the latter. We will not go into it here, though.
\section{Proof, part $(\mathrm b)$}
Here we check that our solution is indeed a general
one. We begin with numerology.

It is easy to figure out that any solution to the
Y-system is completely determined by its values at
$j=-1$ and $j=0$. There are $2r\rp$ of those in the
original formulation~(\ref{orig}), but since we have
eliminated half of the lattice, we only have half
that number left, that is $r\rp$. On the other hand,
the solution we are looking at depends on
$r+\rp+2$ points in $\mathbb{CP}^r$, but due to
projective invariance, $r+2$ of them can be chosen
arbitrarily. This leaves $\rp$ points in an
$r$-dimensional space, which also gives $r\rp$.
So the numbers match up, but now we have to
prove that this is no accident.

First, a lemma.\\[1ex]
\noindent {\bf Lemma.}
\em Two solutions to the Y-system~(\ref{ysys})
coincide everywhere on the lattice
$L=\{(i,j,\?k)\in\mathbb Z^3\?|\?1\leq i\leq r,
1\leq k\leq\rp,i+j+\?k\text{ even}\}$ if they
coincide on either of the following subsets of this
lattice:
\begin{itemize}
\item[\it a)] $J_{-1}\cup J_{\?0}\,$,
\item[\it b)] $Q_0\,$,
\item[\it c)] $K_1\cap
(Q_0\cup Q_1\cup\dotsb\cup Q_{\rp-1})\,$,
\end{itemize}
where $J_n$, $K_n$ and $Q_n$ are intersections
of $L$ with the planes $j=n$, $k=n$ and
$i+j+\?k=-2n$ respectively.
\em

The first item is rather obvious. It has already been
used in the numerological part, and is of no further
use here.

To prove item (b), consider first the equation of the
Y-system~(\ref{ysys}) with $(i,j,\?k)=(1,-3,1)$:
\begin{equation*}
   Y_{1\?-4\?1}=\frac{1+Y_{2\?-3\?1}}
   {Y_{1\?-2\?1}\left(1+1/Y_{1\?-3\?2}\right)}\;.
\end{equation*}
All three lattice points involved in the right hand
side belong to $Q_0$, while the point $(1,-4,1)$ in
the left hand side is the only point of the
intersection $Q_1\cap J_{-4}$. Hence, if two
solutions coincide on $Q_0$, they also coincide on
$Q_1\cap J_{-4}$. But now the next two equations,
\begin{gather*}
   Y_{2\?-5\?1}=\frac{\bigl(1+Y_{3\?-4\?1}\bigr)
   \bigl(1+Y_{1\?-4\?1}\bigr)}
   {Y_{2\?-3\?1}
   \left(1+1/Y_{2\?-4\?2}\right)}\\[1ex]
   Y_{1\?-5\?2}=\frac{1+Y_{2\?-4\?2}}
   {Y_{1\?-3\?2}
   \left(1+1/Y_{1\?-4\?3}\right)
   \left(1+1/Y_{1\?-4\?1}\right)}\;,
\end{gather*}
have the right hand sides entirely contained in
$Q_0\cup(Q_1\cap J_{-4})$ and the left hand sides
spanning $Q_1\cap J_{-5}$. Hence, our solutions
coincide on $Q_1\cap J_{-5}$ as well.

Clearly, this process can be continued step by step
to cover the entire $Q_1$, and then repeated again
and again for all $Q_n$ with $n>0$. That done, we
can start over from $Q_{-1}\cap J_{2-r-\rp}$ and
move in the opposite direction to cover all $Q_n$
with $n<0$ as well. This settles item (b).

On to item (c), we begin with the $r(\rp-1)$
equations
\begin{equation*}
   1+1/Y_{ij\?2}
   =\frac{\left(1+Y_{i+1j\?1}\right)
   \left(1+Y_{i-1j\?1}\right)}
   {Y_{i\,j+1\?1}\?Y_{i\,j-1\?1}}\qquad
   -2\rp+2\leq i+j\leq-2\;.
\end{equation*}
These imply that if two solutions coincide on
$K_1\cap(Q_0\cup Q_1\cup\dotsb\cup Q_{\rp-1})$,
then they also coincide on
$K_2\cap(Q_0\cup Q_1\cup\dotsb\cup Q_{\rp-2})$.
Then the next $r(\rp-2)$ equations,
\begin{equation*}
   1+1/Y_{ij\?3}
   =\frac{\left(1+Y_{i+1j\?2}\right)
   \left(1+Y_{i-1j\?2}\right)}
   {Y_{i\,j+1\?2}\?Y_{i\,j-1\?2}
   \left(1+1/Y_{ij\?1}\right)}\qquad
   -2\rp+3\leq i+j\leq-3\;,
\end{equation*}
give
$K_3\cap(Q_0\cup Q_1\cup\dotsb\cup Q_{\rp-3})$,
and so it continues all the way to
$K_{\rp}\cap Q_0$. But~clearly, the union of
these~$\rp$ sets contains $Q_0$, and item (c) has
thus reduced to item (b). This settles the lemma,
and we return to the main proof.

Fix the points $x_0,\dotsc,x_{r+1}$ in general
position and write out formula~(\ref{yc}) for
$(i,j,\?k)\in
K_1\cap(Q_0\cup Q_1\cup\dotsb\cup Q_{\rp-1})$
as~$\rp$ systems of $r$ equations each for $\rp$
unknown points $x_{r+2},\dotsc,x_{r+\rp+1}$:
\begin{gather*}
   \frac{|\vec x_0\dotsc\vec x_i\,
   \vec x_{i+2}\dotsc\vec x_{r+1}|\?
   |\vec x_1\dotsc\vec x_{i-1}\,
   \vec x_{i+1}\dotsc\vec x_{r+2}|}
   {|\vec x_0\dotsc\vec x_{i-1}\,
   \vec x_{i+2}\dotsc\vec x_{r+2}|\?
   |\vec x_1\dotsc\vec x_{r+1}|}
   =Y_{i,-i-1,1}\\[1ex]
   \frac{|\vec x_1\dotsc\vec x_{i+1}\,
   \vec x_{i+3}\dotsc\vec x_{r+2}|\?
   |\vec x_2\dotsc\vec x_{i}\,
   \vec x_{i+2}\dotsc\vec x_{r+3}|}
   {|\vec x_1\dotsc\vec x_{i}\,
   \vec x_{i+3}\dotsc\vec x_{r+3}|\?
   |\vec x_2\dotsc\vec x_{r+2}|}
   =Y_{i,-i-3,1}\\\vdots\\
   \frac{|\vec x_{\rp-1}\dotsc\vec x_{i+\rp-1}\,
   \vec x_{i+\rp+1}\dotsc\vec x_{r+\rp\?}|\?
   |\vec x_{\rp}\dotsc\vec x_{i+\rp-2}\,
   \vec x_{i+\rp}\dotsc\vec x_{r+\rp+1}|}
   {|\vec x_{\rp-1}\dotsc\vec x_{i+\rp-2}\,
   \vec x_{i+\rp+1}\dotsc\vec x_{r+\rp+1}|\?
   |\vec x_{\rp}\dotsc\vec x_{r+\rp\?}|}
   =Y_{i,-i-2\rp+1,1}\;.
\end{gather*}
Clearly, we shall be done if we show that these
can be solved for any values of the $Y$'s in the
right hand sides, except perhaps $0$, $-1$
and~$\infty$.

Note that the first of the above systems contains
only one unknown point, $x_{r+2}$, and its solution
is easily found to be
\[
   x_{r+2}=t\?(x_0)\;,
\]
where $t$ is the projective transformation whose
linear counterpart $T$ has
$\vec x_1,\dotsc,\vec x_{r+1}$ for eigenvectors,
\[
   T\vec x_n=\lambda_n\vec x_n\qquad\qquad
   n=1,\dotsc,r+1\;,
\]
and the respective eigenvalues are given by
\[
   \lambda_n=\prod_{i=n}^r\,
   \frac1{1+\cfrac1{Y_{i,-i-1,1}}}\;.
\]
It is also easy to show that the newfound point
$x_{r+2}$ is in general position relative to the
points $x_1,\dotsc,x_{r+1}$. This lets us use the
second system in the same way to find $x_{r+3}$,
then use the third one to find $x_{r+4}$ and so on
all the way to the last point $x_{r+\rp+1}$. Thus,
the solution in question is indeed a general one,
and the Theorem is fully proved.

As a final remark, note that the change of indices
$(i,j,\?k)\rightarrow(r+1-i,j+r+\rp+2,\?\rp+1-\?k)$
translates, in terms of the numbers $a\ldots\?d$,
into the pairwise shift/permutation
$(a,b,c,d)\rightarrow(c-r-\rp-2,d-r-\rp-2,a,b)$.
But clearly, the latter has no effect on the general
solution~(\ref{y}), and so our theorem can be
improved somewhat by replacing the periodicity
condition~(\ref{per}) with the stronger one:
\begin{equation*}
   Y_{r+1-i,j+r+\rp+2,\?\rp+1-\?k}=Y_{ij\?k}\;.
\end{equation*}
In fact, Zamolodchikov's conjecture has always been
formulated in this stronger form. The only reason to
start with the weaker version was to slightly
simplify exposition.


\begin{thebibliography}{9}
\bibitem{z} Al. B. Zamolodchikov,
On the thermodynamic Bethe ansatz equations for
reflectionless  ADE scattering theories,
{\it Phys. Lett. B} {\bf 253} (1991), 391--394.
\bibitem{fs} E. Frenkel and A. Szenes,
Thermodynamic Bethe ansatz
and dilogarithm identities. I,
{\it Math. Res. Lett.} {\bf 2} (1995),
no. 6, 677--693.
\bibitem{gt} F. Gliozzi and R. Tateo,
Thermodynamic Bethe ansatz and three-fold
triangulations,
{\it Internat. J. Modern Phys. A} {\bf 11} (1996),
no. 22, 4051--4064.
\bibitem{fz} S. Fomin and A. Zelevinsky,
Y-systems and generalized associahedra,
{\it Ann. of Math.} {\bf 158} (2003),
no. 3 , 977--1018.
\end{thebibliography}
\end{document}